\documentclass[12pt,twocolumn]{iopart}

\usepackage{graphicx,epsfig,cite,subfigure}
\usepackage{bm,subfigure,iopams}

\begin{document}

\title[]{Fluctuation relations for heat engines in time-periodic steady states}

\author{Sourabh Lahiri, Shubhashis Rana and A. M. Jayannavar}

\address{
Institute of Physics, Sachivalaya Marg, Bhubaneswar, Orissa, India - 751005 \eads{\mailto{lahiri@iopb.res.in}, \mailto{shubho@iopb.res.in}, \mailto{jayan@iopb.res.in}}
}

\newcommand{\nwc}{\newcommand}
\nwc{\la}{\langle}
\nwc{\ra}{\rangle}
\nwc{\lw}{\linewidth}
\nwc{\nn}{\nonumber}
\nwc{\Ra}{\Rightarrow}
\nwc{\dg}{\dagger}

\nwc{\pd}[2]{\frac{\partial #1}{\partial #2}}
\nwc{\ppd}[2]{\frac{\partial^2 #1}{\partial #2^2}}

\nwc{\zprl}[3]{Phys. Rev. Lett. ~{\bf #1},~#2~(#3)}
\nwc{\zpre}[3]{Phys. Rev. E ~{\bf #1},~#2~(#3)}
\nwc{\zpra}[3]{Phys. Rev. A ~{\bf #1},~#2~(#3)}
\nwc{\zjsm}[3]{J. Stat. Mech. ~{\bf #1},~#2~(#3)}
\nwc{\zepjb}[3]{Eur. Phys. J. B ~{\bf #1},~#2~(#3)}
\nwc{\zrmp}[3]{Rev. Mod. Phys. ~{\bf #1},~#2~(#3)}
\nwc{\zepl}[3]{Europhys. Lett. ~{\bf #1},~#2~(#3)}
\nwc{\zjsp}[3]{J. Stat. Phys. ~{\bf #1},~#2~(#3)}
\nwc{\zptps}[3]{Prog. Theor. Phys. Suppl. ~{\bf #1},~#2~(#3)}
\nwc{\zpt}[3]{Physics Today ~{\bf #1},~#2~(#3)}
\nwc{\zap}[3]{Adv. Phys. ~{\bf #1},~#2~(#3)}
\nwc{\zjpcm}[3]{J. Phys. Condens. Matter ~{\bf #1},~#2~(#3)}
\nwc{\zjpa}[3]{J. Phys. A: Math. Theor. ~{\bf #1},~#2~(#3)}
\nwc{\zpjp}[3]{Pram. J. Phys. ~{\bf #1},~#2~(#3)}
\nwc{\zpa}[3]{Physica A ~{\bf #1},~#2~(#3)}

\begin{abstract}
 A fluctuation relation for  heat engines (FRHE) has been derived recently. In the beginning, the system is in contact with the cooler bath. The system is then coupled to the hotter bath and external parameters are changed cyclically, eventually bringing the system back to its initial state, once the coupling with the hot bath is switched off. In this work, we lift the condition of initial thermal equilibrium and derive a new fluctuation relation for the central system (heat engine) being in a time-periodic steady state (TPSS).  Carnot's inequality for classical thermodynamics follows as a direct consequence of this fluctuation theorem even in TPSS. For the special cases of the absence of hot bath and no extraction of work, we obtain the integral fluctuation theorem for total entropy and the generalized exchange fluctuation theorem, respectively. Recently microsized heat engines have been realized experimentally in the TPSS. We numerically simulate the same model and verify our proposed theorems.
\end{abstract}



\maketitle


\section{Introduction}
Over the past two decades several exact and unexpected relations for exchange of energy, heat, entropy, etc. have been obtained that remain valid even for systems driven far away from thermal equilibrium.  These results, although in principle valid quite generally, are in practice relevant mostly for microscopic systems for which the fluctuations are substantial.  These are called \emph{fluctuation theorems}\cite{eva93,eva94,jar97,cro98,cro99}. They transform classical thermodynamic inequalities into equalities. Advances in experimental techniques have made dramatic progress in the area of single-molecule manipulation and nanotechnology have led to experimental verification of the various fluctuation theorems  \cite{pit11,rit06}. 
Recently another equality is added to the class of fluctuation theorems, namely, fluctuation relations for heat engines (FRHE) \cite{sin11}.   Initially the system is in thermal equilibrium with a cold thermal reservoir at temperature $T_c$, and then coupled to a hot thermal reservoir at temperature $T_h> T_c$. At this stage, the parameters driving the working substance (our system of interest) are changed cyclically so that at the end of the cycle all the parameters attain their initial values, and the interaction with the hot reservoir is switched off, and the system is coupled to the cold reservoir. The equality reads
\begin{equation}
  \left< \exp\left[-Q_h\left(\frac{1}{T_c}-\frac{1}{T_h}\right)+\frac{W}{T_c}\right]\right> = 1.
  \label{FRHE}
\end{equation}
Here, $\la\cdots\ra$ denotes averaging over many realizations of the cycle. $Q_h$ is the heat \emph{absorbed} from the hot bath and $W$ is the work \emph{extracted} from the system in a cycle. 

\section{Derivation from Seifert's theorem}
We provide here a derivation slightly different from that given in \cite{sin11,rit06}, by using the 
Seifert's integral fluctuation theorem \cite{sei05,sei08}, $\la e^{-\Delta s_{tot}}\ra = 1$ (in presence of multiple baths) in conjunction with the first law, 
\begin{equation}
  \Delta E = Q_h-Q_c-W
\end{equation}
with the total entropy being given by
\begin{equation}
  \Delta s_{tot} = \Delta s_h+\Delta s_c + \Delta s.
  \label{law1}
\end{equation}
$\Delta s_h$, $\Delta s_c$ and $\Delta s$ are the entropy changes of the hot bath, the cold bath and of the central system, respectively. Denoting the initial and final distributions for the forward process by $p_0(x_0)$ and $p_1(x_\tau)$, we have \cite{sei05,sei08}
\begin{eqnarray}
  \Delta s_h &=& -\frac{Q_h}{T_h}; \hspace{1cm} \Delta s_c = \frac{Q_c}{T_c}; \\
  \Delta s &=& \ln \frac{p_0(x_0)}{p_1(x_\tau)} = \ln \left[\frac{e^{-E(x_0)/T_c}}{Z_0}\cdot \frac{Z_\tau}{e^{-\beta E(x_\tau)/T_c}}\right] = \frac{\Delta E}{T_c},
\end{eqnarray}
where $\Delta E = E(x_\tau)-E(x_0)$, and we have made use of the fact that for a cyclic process, $Z_0=Z_\tau$. Using the first law, eq. (\ref{law1}), we have $Q_c = Q_h-W-\Delta E$. Thus, $\Delta s_{tot}$ becomes 
\begin{eqnarray}
  \Delta s_{tot} &=& -\frac{Q_h}{T_h} + \frac{Q_c}{T_c} + \frac{\Delta E}{T_c} \nn\\
  &=& -\frac{Q_h}{T_h} + \frac{Q_h-W-\Delta E}{T_c} + \frac{\Delta E}{T_c} \nn\\
  &=& Q_h\left(\frac{1}{T_c}-\frac{1}{T_h}\right) - \frac{W}{T_c}.
\end{eqnarray}
Seifert's theorem then gives eq. (\ref{FRHE}).

Equality (\ref{FRHE}), together with the Jensen's inequality gives
\begin{equation}
  \left<Q_h\left(\frac{1}{T_c}-\frac{1}{T_h}\right)-\frac{W}{T_c}\right> \ge 0,
\end{equation}
which can be rewritten as
\begin{equation}
  \frac{\la W\ra}{\la Q_h\ra} \le \eta_c,
\end{equation}
$\eta_c$ being the Carnot efficiency given by $\eta_c\equiv 1-\frac{T_c}{T_h}$. This is then, the Carnot's theorem for maximum efficiency applied to a mesoscopic heat engine. Now, instead of taking the averaged quantities, one can also define the efficiency for each individual trajectory, $\eta\equiv W/Q$, which is, of course, a fluctuating quantity. Consequently, there may be trajectories along which $\eta>\eta_c$, which will be termed as the {\it atypical trajectories} (trajectories that seem to flout the behaviour dictated by the second law). In fact, $\eta$ can also become negative, in which case, along a cycle, the system does not perform as a heat engine (for example, when heat is absorbed by the system, but work is being done \emph{on} the system \cite{ran}).

\section{Derivation of the FRHE for a time-periodic steady state}
Recently the Carnot engine has been investigated experimentally in the time-periodic steady state (TPSS) \cite{bli12}. In a TPSS, the probability density of system state, $p_{ss}(x,t)$, is periodic in time, $p_{ss}(x,t+\tau)=p_{ss}(x,t)$, where $\tau$ is the time-period of the external drive. 
The occupation probabilities of a motor in a TPSS, consisting of a two-level system, has been studied in \cite{chv10}.
In a TPSS, the probability density for the system state can be written as $p_{ss}(x,\lambda)=e^{-\phi(x,\lambda)}$, $\lambda$ being the external time-dependent protocol, and $\phi(x,\lambda)$ is an effective potential. In such a case, the condition of initial equilibration of the working substance with the cold bath ought to be lifted. Once again, in a part of the cycle, the system is connected to the cold bath, while in the other part, it is connected to the hot bath. In this case, the change in system  entropy during a cycle is given by $\Delta \phi$, and the change in the total entropy becomes
\begin{eqnarray}
  \Delta s_{tot} &=& -\frac{Q_h}{T_h} + \frac{Q_c}{T_c} + \Delta \phi \nn\\
  &=& -\frac{Q_h}{T_h} + \frac{Q_h-W-\Delta E}{T_c} + \Delta \phi \nn\\
  &=& Q_h\left(\frac{1}{T_c}-\frac{1}{T_h}\right) - \frac{W+\Delta E}{T_c}+\Delta \phi.
\label{stot}
\end{eqnarray}
Let $X$ denote the short form for a trajectory in phase space: $\{x_0\to x_1 \to x_2 \to\cdots\to x_\tau\}$, and let $\tilde X$ denote the time-reversed path: $\{x_0\leftarrow x_1 \leftarrow x_2 \leftarrow\cdots\leftarrow x_\tau\}$, the subscripts denoting discretized time. According to the detailed fluctuation theorem for total entropy \cite{sei05,sei08}, 
we then have the following ratio between the probability densities for the forward and reverse trajectories, represented by $P[X]$ and $\tilde P[\tilde X]$, respectively:
\begin{equation}
  \frac{P[X]}{\tilde P[\tilde X]} = e^{\Delta s_{tot}} = \exp\left[Q_h\left(\frac{1}{T_c}-\frac{1}{T_h}\right)-\frac{W+\Delta E}{T_c}+\Delta \phi\right],
  \label{FRHE1}
\end{equation}
whose integrated form is given by the new equality
\begin{equation}
  \left<\exp\left[-Q_h\left(\frac{1}{T_c}-\frac{1}{T_h}\right)+\frac{W+\Delta E}{T_c}-\Delta \phi\right]\right>=1.
  \label{FRHE2}
\end{equation}
{\bf We next derive a detailed fluctuation theorem for the joint probability distribution for work, heat, change in internal energy and system entropy}.

\section{Fluctuation theorem for the joint probability distribution}

Using eq. (\ref{FRHE1}), we  obtain a relation for the joint probability density for $Q_h$, $W$, $\Delta E$ and $\Delta\phi$. These quantities are odd under time-reversal.
\begin{eqnarray}
  &&P(Q_h,W,\Delta E,\Delta\phi) = \int \mathcal D[X] ~P[X]~\delta(Q_h-Q_h[X])~\delta(W-W[X])~\nn\\
  &&\hspace{1cm}\times \delta(\Delta E-\Delta E(x_0,x_\tau))~\delta(\Delta\phi-\Delta\phi(x_0,x_\tau))\nn\\
  &&\nn\\
  &=& \int \mathcal D[X] ~P[\tilde X]\exp\left[-Q_h\left(\frac{1}{T_c}-\frac{1}{T_h}\right)+\frac{W+\Delta E}{T_c}-\Delta \phi\right]~\delta(Q_h-Q_h[X])\nn\\
  &&\hspace{1cm}\times\delta(W-W[X])~\delta(\Delta E-\Delta E(x_0,x_\tau))~\delta(\Delta\phi-\Delta\phi(x_0,x_\tau)). \nn
\end{eqnarray}
Here, $\mathcal D[X] =  \mathcal D[\tilde X] = dx_0dx_1\cdots dx_\tau$, where $\tilde x$ is the time-reversed state of $x$.
We now now perform a change of variables from $x$ to $\tilde x$. Then,
\begin{eqnarray}
  P(Q_h,W,\Delta E,\Delta\phi) &=& \exp\left[Q_h\left(\frac{1}{T_c}-\frac{1}{T_h}\right)-\frac{W+\Delta E}{T_c}+\Delta \phi\right] \nn\\
  &&\times\int \mathcal D[X]~\delta(Q_h+\tilde Q_h[\tilde X])~\delta(W+\tilde W[\tilde X])\nn\\
  &&\hspace{1cm}\times~\delta(\Delta E+\Delta\tilde E(\tilde x_0,\tilde x_\tau))~\delta(\Delta\phi+\Delta\tilde\phi(\tilde x_0,\tilde x_\tau)) \nn\\
\nn\\
  &=& \tilde P(-Q_h,-W,-\Delta E,-\Delta\phi)\nn\\
  &&\times~\exp\left[Q_h\left(\frac{1}{T_c}-\frac{1}{T_h}\right)-\frac{W+\Delta E}{T_c}+\Delta \phi\right].
\end{eqnarray}
Here, $\tilde P(-Q_h,-W,-\Delta E,-\Delta\phi)$ is the joint probability density for $-Q_h$, $-W$, $-\Delta E$ and $-\Delta \phi$, along the reverse process. Noting that in a TPSS, $P$ and $\tilde P$ have the same functional forms, we can write
\begin{equation}
  \frac{P(Q_h,W,\Delta E,\Delta\phi)}{P(-Q_h,-W,-\Delta E,-\Delta\phi)} = \exp\left[Q_h\left(\frac{1}{T_c}-\frac{1}{T_h}\right)-\frac{W+\Delta E}{T_c}+\Delta \phi\right].
  \label{FRHE3}
\end{equation}
Eq. (\ref{FRHE3}) readily leads to eq. (\ref{FRHE2}), which in turn gives rise to the inequality
\begin{eqnarray}
  &\la Q_h\ra \left(\frac{1}{T_c}-\frac{1}{T_h}\right) - \frac{\la W\ra+\la\Delta E\ra-T_c\la\Delta \phi\ra}{T_c} \ge 0\nn\\
  \label{ineq1}
\end{eqnarray}
In the TPSS, we have, $\la\Delta E\ra = 0$, and $\la\Delta \phi\ra = 0$. Then we arrive, even for the TPSS, to the Carnot's theorem, namely,
\begin{equation}
  \frac{\la W\ra}{\la Q_h\ra} \le \eta_c.
\end{equation}

For $T_c=T_h$ (system is in contact with a single bath), we retrieve the Seifert's integral fluctuation theorem from eq. (\ref{FRHE2}) for a system in contact with a bath at temperature $T_c$ \cite{sei05,sei08}:
\begin{eqnarray}
  \left<\exp\left[\frac{W+\Delta E}{T_c}-\Delta \phi\right]\right> =  \left<\exp\left[ -\frac{Q_c}{T_c}-\Delta \phi\right]\right> = \la e^{-\Delta s_{tot}}\ra = 1.
\end{eqnarray}
We have used the first law for system in contact with only the cold bath, $\Delta E = -W-Q_c$, the first step. 
If no work is extracted from the system, then the system effectively acts as a heat conductor between the two heat baths, giving rise to the generalized exchange fluctuation theorem \cite{jar04} in TPSS:
\begin{equation}
  \left<\exp\left[-Q_h\left(\frac{1}{T_c}-\frac{1}{T_h}\right)+\frac{\Delta E}{T_c}-\Delta \phi\right]\right> = 1.
\label{XFT}
\end{equation}
An example of the above case (eq. (\ref{XFT})) would be a particle in a harmonic potential coupled to a bath whose temperature changes periodically in time, while  no other parameters of the harmonic oscillator are changed, and consequently work extracted is zero. This model should be experimentally realizable.

To verify our proposed theorem, eq. (\ref{FRHE2}), we study a simple heat engine which has been experimentally realized recently. Some related points have been clarified through the simulations of the distribution functions of physical quantities appearing in our theorems.

\section{The model and numerical results}
\label{sec:num}
In this section, we verify eq. (\ref{FRHE2}) numerically. For this purpose, we choose the model used in \cite{bli12}, namely, the mesoscopic realization of a Stirling engine. Each cycle in its operation consists of the following steps:
\begin{enumerate}

\item {\bf Step 1}: an overdamped colloidal particle is initially trapped in a harmonic potential with a spring constant $k_{min}$ (state $A$): $V(x,0) = \frac{1}{2}k_{min}x^2$. The particle is in contact with a medium of temperature $T_c$. Without breaking contact with the heat bath, the stiffness constant is subsequently changed, via a prescribed time-dependence of this constant $k(t)$, until it reaches a value $k_{max}$ (state $B$) after a time $\tau$. The potential function now is given by $V(x,\tau) = \frac{1}{2}k_{max}x^2$.

\item {\bf Step 2}: the bath temperature is suddenly switched to $T_h>T_c$ (state $C$). The distribution of states of the system does not change during this instantaneous jump.

\item {\bf Step 3}: now the spring constant follows a separate time dependence due to which its value changes from $k_{max}$ to $k_{min}$ (state $D$) over a time $\tau$.

\item {\bf Step 4}: in the last step, the temperature of the medium is once again instantaneously switched back to its initial value $T_c$ and the system returns to state $A$. The full cycle is then repeated.
\end{enumerate}
Since in steps 2 and 4, the stiffness constant is held fixed, the work done is identically equal to zero in these two steps. 
We choose the functional dependence for the stiffness constant during the transition  state $A$ $\to$ state $B$ to be linear and of the following form:
\begin{equation}
k(t) = k_{min}+q\left(\frac{t}{\tau}\right).
\end{equation}
According to this equation, after time $\tau$, the system reaches $k_{max} = k(\tau) = k_{min}+q$. Similarly, during the transition state $C$ $\to$ state $D$, the form of $k(t)$ is given by
\begin{equation}
k(t) = k_{max} - q\left(\frac{t}{\tau}-1\right).
\end{equation}
We find that when the full cycle is complete, i.e. $t=2\tau$, we get back the initial spring constant $k_{min} = k(2\tau) = k_{max}-q$. In our simulation, we choose the values of the constants (in dimensionless units) to be $k_{min}=1$, $k_{max}=2$, $T_c=0.1$ and $T_h=0.4$. Initially, as a consistency check, we verify that for a very slow process (time of observation large compared to the relaxation period  of the system to its equilibrium state), the average work done on the system equals the change in its free energy (quasi-static process). 
In our simulation, we have used Heun's method of integration and have generated $\sim 10^5$ state space trajectories.
The changes in free energy during the steps 1 and 3 are
\begin{equation}
\Delta F_{A\to B} = \frac{T_c}{2}\ln\frac{k_{max}}{k_{min}}
\end{equation}
and
\begin{equation}
\Delta F_{C\to D} = \frac{T_h}{2}\ln\frac{k_{min}}{k_{max}},
\end{equation}
respectively. For our chosen parameters, we get $\Delta F_{1\to 2} = 0.035$ and $\Delta F_{3\to 4} = -0.139$. From our simulation, we obtain the average works done in steps 1 and 3 reach these values as we increase the time of observation. For $\tau=50$, we obtain $\la W\ra_{A\to B}= 0.036$ and $\la W\ra_{C\to D}=-0.138$, respectively, which match with the theoretical results, within our numerical accuracy. 

For reaching the time-periodic steady state, we leave out several initial cycles to skip the transient regime. For this TPSS, we have chosen $\tau=5$, and we obtain the value of eq. (\ref{FRHE2}) to be 1.083, which is very close to unity. Thus, the above relation is verified in our numerical simulations. 

\begin{figure}[!h]
\centering
\subfigure[]{\includegraphics[width=0.45\lw]{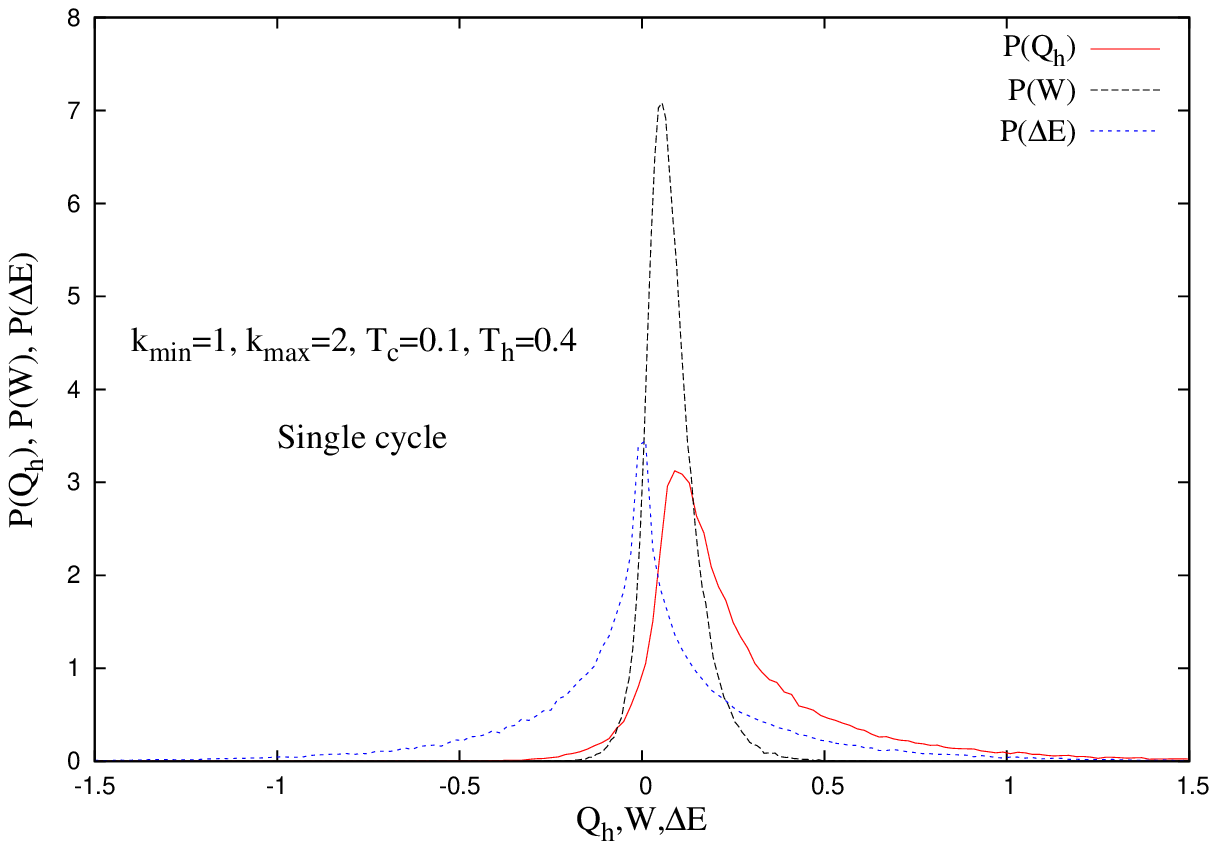}}
\subfigure[]{\includegraphics[width=0.45\lw]{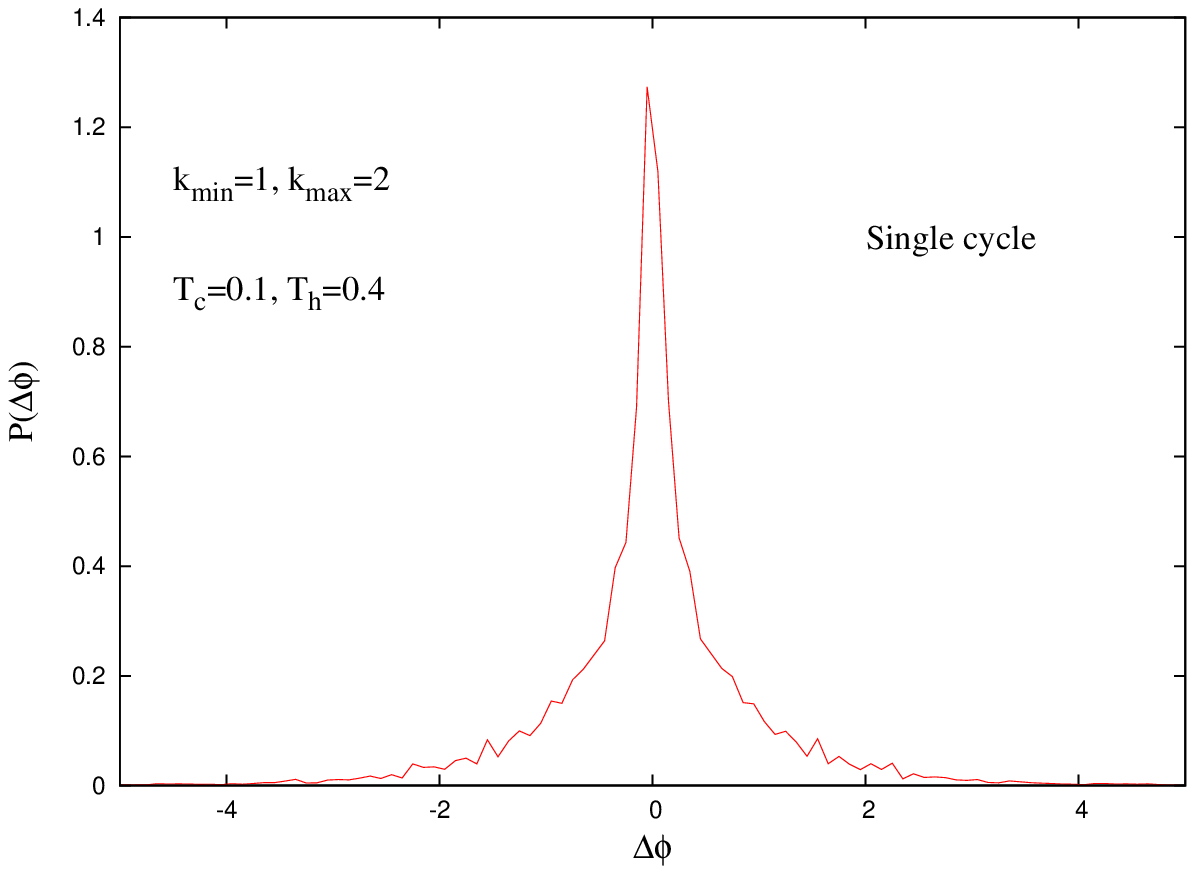}}
\caption{(a) Distribution of $Q_h$, $W$ and $\Delta E$ for a single cycle of the heat engine in a steady state, for $\tau=5$. (b) Distribution of $\Delta \phi$ for the same parameters.}
\label{fig:1cycle}
\end{figure}

Now we study the behaviour of $\Delta s_{tot}$ (eq. (\ref{stot})) when each realization of the experiment consists of a large number of cycles. It apparently seems that since $\Delta E$ and $\Delta\phi$ are state functions, while $Q_h$ and $W$ scale with time of observation, in the limit of a large number of cycles, we will have vanishing contribution from the state functions to the fluctuation theorem. To verify this numerically, 
in figure \ref{fig:1cycle}(a), we have plotted the distributions for $Q_h$, $W$ and $\Delta E$ for a single cycle of the heat engine. $\Delta E$ being a state function is symmetric about the $\Delta E=0$ axis. In figure \ref{fig:1cycle}(b), the distribution for change in system entropy, $\Delta \phi$, is plotted for a single cycle.

\begin{figure}[!h]
\centering
\includegraphics[width=0.5\lw]{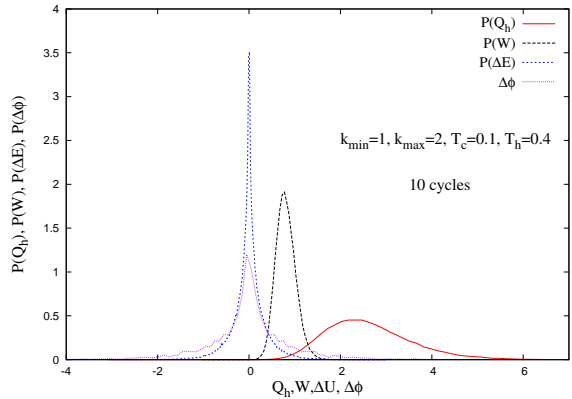}
\caption{Distribution of $Q_h$, $W$ and $\Delta E$ for 10 cycles of the heat engine in a steady state, with the half observation time $\tau=5$ for each cycle.}
\label{fig:ncycles}
\end{figure}

In figure \ref{fig:ncycles}, we have plotted the distribution functions for $Q_h$, $W$,  $\Delta E$ and $\Delta \phi$ for 10 cycles in the steady state. As expected, we find that the distributions for $Q_h$ and $W$ tend towards a Gaussian  and shift towards right, but those for $\Delta E$  and $\Delta \phi$ remain similar to the case of a single cycle. 

\begin{figure}[!h]
\centering
\subfigure[]{\includegraphics[width=0.45\lw]{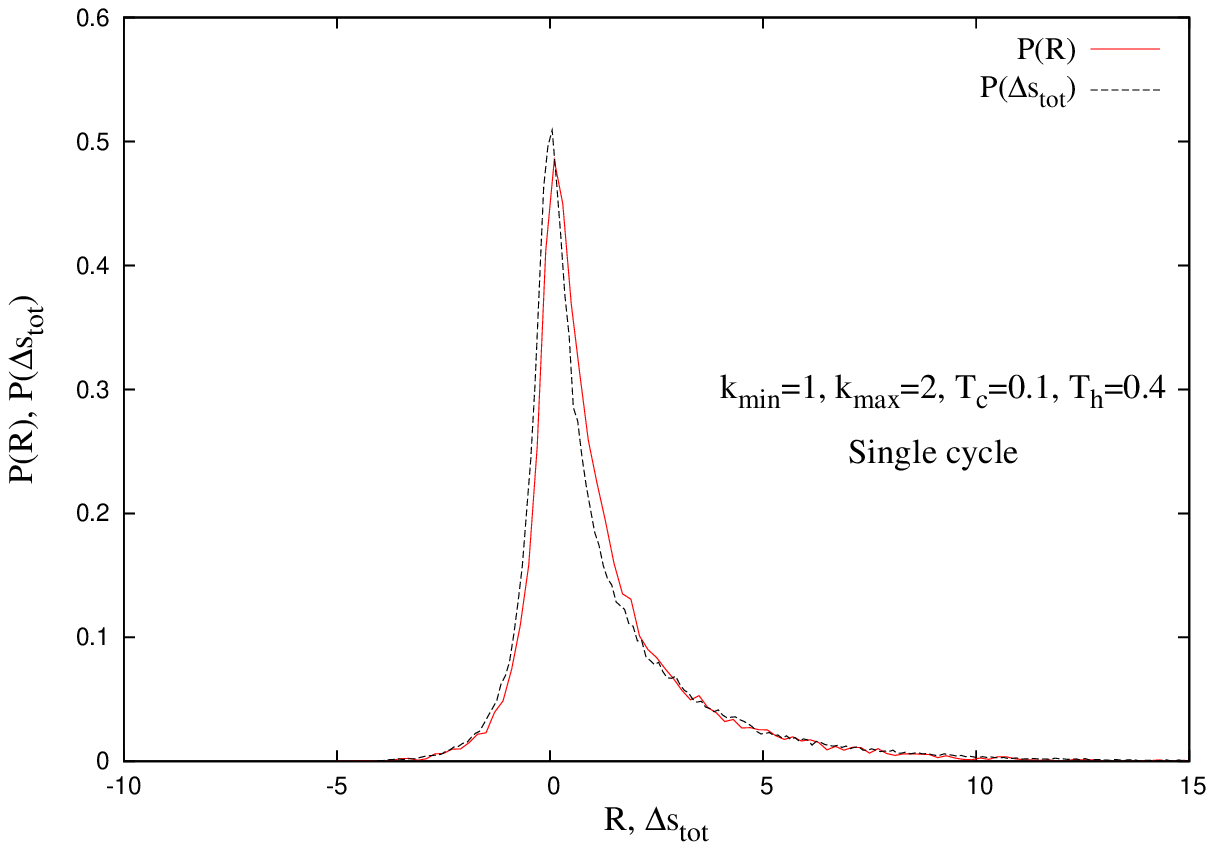}}
\subfigure[]{\includegraphics[width=0.45\lw]{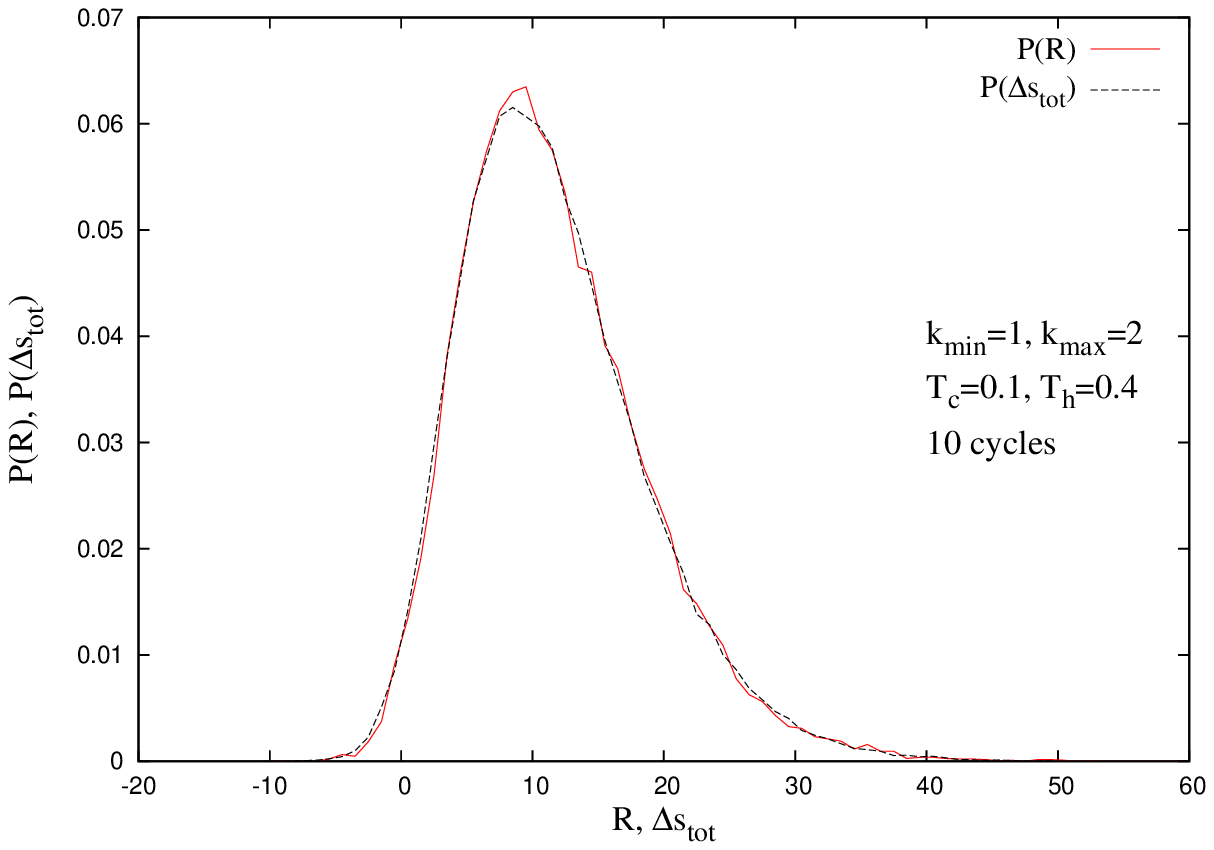}}
\caption{(a) Distribution of the $R\equiv Q_h(1/T_c-1/T_h)-W/T_c$, which is the extensive part of $\Delta s_{tot}$ and of $\Delta s_{tot}$ itself, for a single cycle in steady state. (b) Same distributions for 10 cycles, and we still find an appreciable difference between the two.}
\label{fig:fulldiss}
\end{figure}

In figure \ref{fig:fulldiss}, we have plotted distributions of $R\equiv Q_h(1/T_c-1/T_h)-W/T_c$ (which is the extensive part of $\Delta s_{tot}$) and of $\Delta s_{tot}$ itself. In figure \ref{fig:fulldiss}(a), we find that the two quantities follow distributions that are slightly different from each other. In figure \ref{fig:fulldiss} (b), we find that when we take a large number of cycles, the distributions begin to coincide. This is because the contribution from the distributions of state functions become small as compared to the contributions from the extensive quantities in the limit of large number of cycles. However, it may be noted that the intensive quantities do contribute in the extreme tails of the distributions (large deviation). To see this contribution we need very high precision simulation in the tail region, which is beyond the accuracy of our simulation. This point also arises in the case of heat and work theorems. Work obeys a fluctuation theorem. However, due to the contribution from the internal energy change, heat does not follow a fluctuation theorem, even in the limit of large observation time \cite{zon03,zon04}.

\section{Conclusion}

In conclusion, we have generalized the fluctuation relation for heat engines to time-periodic steady states, which leads to the Carnot's theorem. Generalized FRHE leads to, in different limits, to the Seifert's theorem, and the generalized exchange fluctuation theorem. Our FRHE has been verified numerically in a simple realistic heat engine. It would be interesting to check whether the steady state distribution $p_{ss}(x,t)$ in the special case specified below eq. (\ref{XFT}) can be calculated analytically, for example, by generalizing the method given for time-independent steady state in \cite{jar99}. Also, the work distribution for a system starting from equilibrium and trapped in a harmonic potential of time-dependent stiffness constant has been studied in \cite{spe11}. It would be interesting to see whether this procedure can be generalized to deduce the steady state distributions of different thermodynamic quantities for the heat engine considered in section \ref{sec:num}.

\section{Acknowledgement}

One of us (AMJ) thanks DST, India for financial support.

\end{document}